\documentclass[aps,prl,twocolumn,superscriptaddress]{revtex4}
\usepackage{float}
\usepackage{makeidx}
\usepackage{graphicx,amsmath}
\usepackage{CJK}
\usepackage{colordvi}
\usepackage{color}
\usepackage{amssymb}
\usepackage{hyperref}
\usepackage{graphicx}
\usepackage{dcolumn}
\usepackage{bm,amsmath,verbatim}
\usepackage{mathrsfs}
\usepackage{color}
\usepackage[latin9]{inputenc}
\usepackage{amsmath}
\usepackage{amssymb}
\usepackage{graphicx}
\usepackage{esint}
\usepackage{amsfonts}
\usepackage{eurosym}
\usepackage{color}
\usepackage{xcolor}
\usepackage{mathrsfs}
\usepackage{amsmath}
\usepackage{graphicx}
\usepackage{dcolumn}
\usepackage{bm}
\usepackage{times}
\usepackage{amssymb}
\usepackage{float}
\usepackage{array}
\usepackage{tikz}

\def\be{\begin{equation}}
\def\ee{\end{equation}}
\def\bea{\begin{eqnarray}}
\def\eea{\end{eqnarray}}
\def\bse{\begin{subequations}}
\def\ese{\end{subequations}}

\def\be{\begin{eqnarray}}
\def\ee{\end{eqnarray}}

\begin{document}

\title{Synthetic photonic lattices: new routes towards all-optical photonic
devices}
\author{Xi-Wang Luo}
\author{Xingxiang Zhou}
\thanks{email: xizhou@ustc.edu.cn}
\author{Jin-Shi Xu}
\author{Chuan-Feng Li}
\author{Guang-Can Guo}
\affiliation{Key Laboratory of Quantum Information, University of Science and Technology
of China, Hefei, Anhui 230026, China}
\affiliation{Synergetic Innovation Center of Quantum Information and Quantum Physics,
University of Science and Technology of China, Hefei, Anhui 230026, China }
\author{Chuanwei Zhang}
\affiliation{Department of Physics, The University of Texas at Dallas, Richardson, Texas
75080, USA}
\author{Zheng-Wei Zhou}
\thanks{email: zwzhou@ustc.edu.cn}
\affiliation{Key Laboratory of Quantum Information, University of Science and Technology
of China, Hefei, Anhui 230026, China}
\affiliation{Synergetic Innovation Center of Quantum Information and Quantum Physics,
University of Science and Technology of China, Hefei, Anhui 230026, China }

\begin{abstract}
All-optical photonic devices are crucial for many important photonic
technology and applications, ranging from optical communication to quantum
information processing. Conventional design of all-optical devices is based
on photon propagation and interference in real space, which may reply on
large numbers of optical elements and are challenging for precise control.
Here we propose a new route for engineering all-optical devices using photon
internal degrees of freedom, which form photonic crystals in such synthetic
dimensions for photon propagation and interference. We demonstrate this new
design concept by showing how important optical devices such as quantum
memory and optical filter can be realized using synthetic orbital angular
momentum (OAM) lattices in a single main degenerate cavity. The new
designing route utilizing synthetic photonic lattices may significantly
reduce the requirement for numerous optical elements and their fine tuning
in conventional design, paving the way for realistic all-optical photonic
devices with novel functionalities.
\end{abstract}

\maketitle

\section{Introduction}

\begin{figure}[b]
\includegraphics[width=1.0\linewidth]{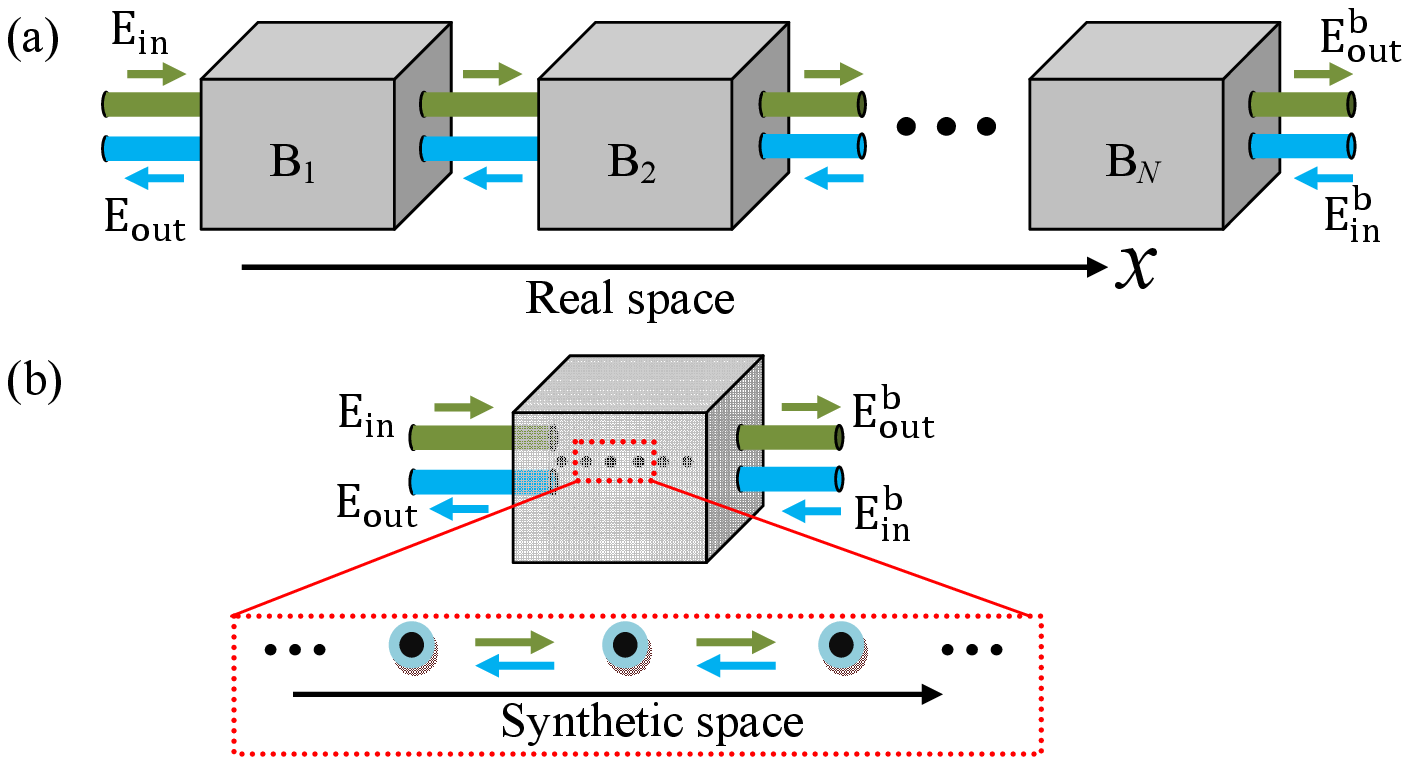}
\caption{\textbf{Illustration of the designing principle for all-optical
photonic devices based on photon interference in synthetic photonic lattices.%
} (a) Conventional devices such as photonic crystals based on photon
interference in real space. The building blocks (boxes $B_{1},B_{2},\ldots $%
), which consist of optical elements such as interferometers or resonators,
are separated in real space and coupled by fibers or waveguides. (b) New
device based on photon interference in synthetic photonic lattices formed by
photon internal degrees of freedom. The dots (lattice sites) represent
different internal states.}
\label{fig:synthetic}
\end{figure}

The ability of coherently controlling the properties of photons, such as
their storage and propagation, is crucial for many important technological
applications in various fields, ranging from optical communications \cite{thevenaz2008slow, vlasov2008high, takesue2013chip}, data storage \cite{yanik2004stopping2, baba2008slow, lvovsky2009optical, kuramochi2014large}, to quantum information
processing \cite{kimble2008quantum, northup2014quantum}. The devices used for such purpose may involve the interaction of
photons with other physical media (\textit{e.g.}, atoms) \cite%
{lvovsky2009optical, northup2014quantum} or contain only optical elements, \textit{%
i.e.}, all-optical photonic devices \cite{yanik2004stopping2, baba2008slow, vlasov2008high, takesue2013chip, kuramochi2014large}. In conventional all-optical devices, photonic properties are controlled through manipulating the photon interference in the real space \cite{baba2008slow, takesue2013chip}. Typical examples include photonic crystals, where coupled arrays of photonic circuits are implemented by fine
tuning the parameters of associated optical elements. The conventional
all-optical photonic devices have been extensively studied and showcase
great applications, ranging from practical devices \cite{baba2008slow, vlasov2008high, kuramochi2014large} to fundamental topological photonics \cite{hafezi2011robust, fang2012realizing, lu2014topological}. However, such real space photonic devices
usually demand precise control of a large number ($\sim 100$) of spatially
separated optical elements (e.g., resonators, waveguides, \textit{etc}.)
[see illustration in Fig. \ref{fig:synthetic} (a)], which can be very
complicated and resource-costing for many practical applications.

On the other hand, photons possess many internal degrees of freedom (e.g.,
frequency, polarization, orbital angular momentum (OAM) \cite%
{yariv2007photonics, allen1992orbital}, \textit{etc}.), which may form
synthetic lattice dimensions for photons (i.e., synthetic photonic crystals)
in addition to real space. Recently synthetic lattice dimensions have been
explored in ultra-cold atomic gases with the direct experimental observation
of quantum Hall edge states \cite{PhysRevLett.112.043001,
mancini2015observation, stuhl2015visualizing}. Here we propose a
conceptually new route for engineering all-optical photonic devices based on
photon propagation and interference in such synthetic photonic lattices [see
Fig. \ref{fig:synthetic} (b)], instead of a large number of optical elements
in the real space photonic crystals. The all-optical devices based on such
synthetic photonic lattices may significantly reduce the physical complexity
of the system and are more resource-efficient.

In this article, we explore this new designing paradigm by showing how
all-optical photonic devices can be implemented using photon propagation and
interference in the synthetic lattices formed by photon OAM modes. Because
of the large number of available distinctive OAM states, photon OAM has
found great applications in quantum information \cite{fickler2014interface,
wang2015quantum, malik2016multi}, optical communications \cite%
{barreiro2008beating, wang2012terabit}, and the realization of topological
matter \cite{luo2015quantum}. Different discrete OAM states form a natural
synthetic lattice for photon interference. We consider a degenerate-cavity
system that supports multiple degenerate OAM modes, where the interference
of photon in the OAM lattice can be manipulated by simply tuning an optical
phase. We show that important photonic devices such as quantum memory and
optical filter that are vital for quantum communication networking and
optical signal processing can be realized using an OAM-based system with
only a single main cavity, where the photon is stopped, stored, and read out
on demand in the synthetic OAM lattice. The proposed all-optical quantum
memory are more resource-efficient and experimentally simpler than
conventional real-space coupled-cavity-based memory that requires precise
control of a large number of coupled cavities \cite{yanik2004stopping}. They
have a large bandwidth and no restriction on the working frequency compared
to atomic-ensemble-based memory \cite{simon2010quantum}. The proposed route
will not only motivate other novel applications and devices based on the OAM
lattices, but also open a completely new avenue for engineering all-optical
photonic devices utilizing other internal degrees of freedom (e.g.,
frequency, etc.) as synthetic lattice dimensions \cite{ozawa2016synthetic,
yuan2016photonic, ozawa2016synthetic2}.

\section{Results}

\subsection{Photon OAM and coupled-degenerate-cavities}

Solutions of the light field in an optical system with cylindrical symmetry
have an angular dependence $e^{il\varphi }$, where $\varphi $ is the
azimuthal angle and $l$ is an integer \cite{yariv2007photonics}. This is a
fundamental optical degree of freedom associated with the OAM of photons
that has a value of $l\hbar $ per photon \cite{allen1992orbital}. In
comparison with other optical degrees of freedom, OAM has a fascinating
property that an infinite number of distinctive OAM states are available.
These discrete OAM $l$ states can be used to denote discrete lattice sites
in the OAM-enabled synthetic lattice dimension.

\begin{figure}[tbp]
\includegraphics[width=1.0\linewidth]{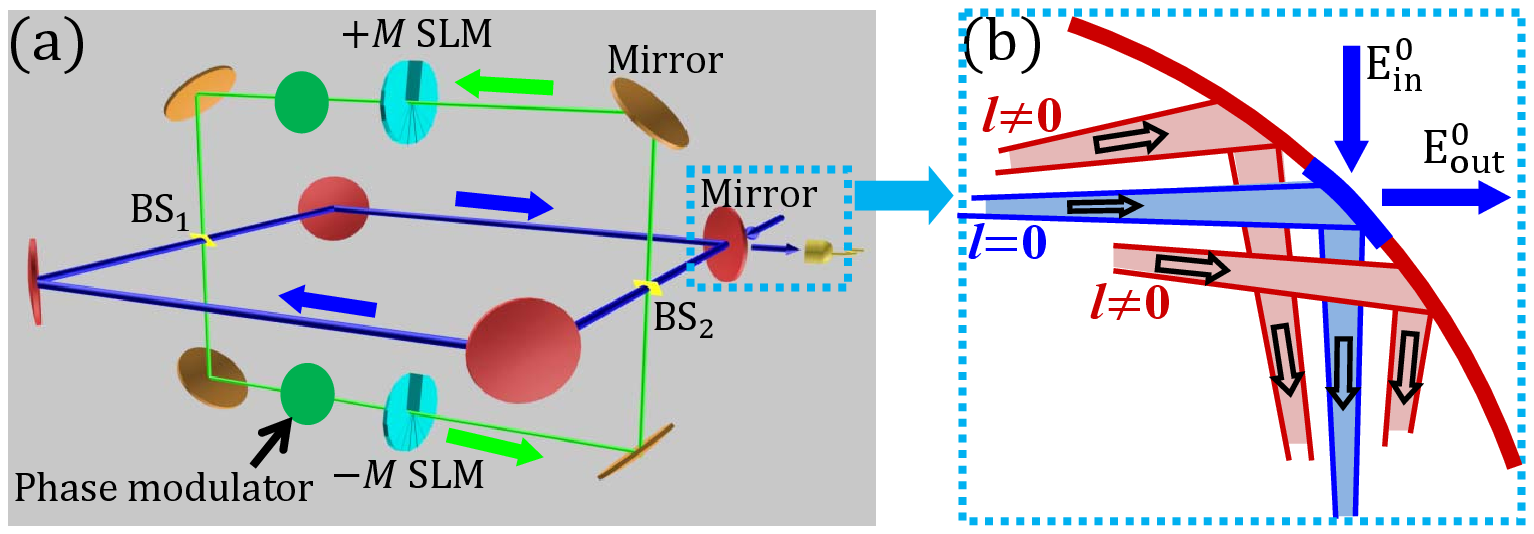}
\caption{\textbf{All-optical photonic devices based on the OAM of photons in
a degenerate cavity system.} (a) The main (red) and auxiliary (brown)
cavities are coupled by a pair of beam splitters, both degenerate. (b) The
input/output port realized as a pinhole at the center of a mirror to couple
the $l=0$ mode in and out. }
\label{fig:system-1}
\end{figure}

The OAM-based synthetic photonic lattices can be generated and manipulated
using a degenerate cavity \cite{arnaud1969degenerate, chalopin2010frequency}
that can support multiple OAM modes simultaneously. In experiments, such a
degenerate cavity with a large number of OAM modes is easy to realize with a
flexible configuration \cite{arnaud1969degenerate}. The proposed all-optical
photonic devices rely on the coupling between different OAM modes in the
degenerate cavity for photon interference and propagation, which can be
realized using an auxiliary cavity. The optical design, shown in Fig. \ref%
{fig:system-1}(a), consists of a main degenerate cavity and an auxiliary
degenerate cavity coupled by two beam splitters with low reflectivity.
Unlike the main cavity, the length of the auxiliary cavity is chosen for
destructive interference, therefore most photons remain in the main cavity.
Two spatial light modulators (SLMs) such as very low-loss vortex phase
plates \cite{oemrawsingh2005experimental} are inserted into the auxiliary
cavity, which couple OAM mode $l$ of the passing photons to its adjacent
modes $l\pm M$ with with $M$ being the step index of the SLM. Two phase
modulators placed in the two arms of the auxiliary cavity generate different
phases for the coupling along two arms, which can be realized using, for
instance, high-speed electro-optic index modulation \cite{chuang1995physics,
yariv2007photonics}.

\begin{figure}[tbp]
\includegraphics[width=1.0\linewidth]{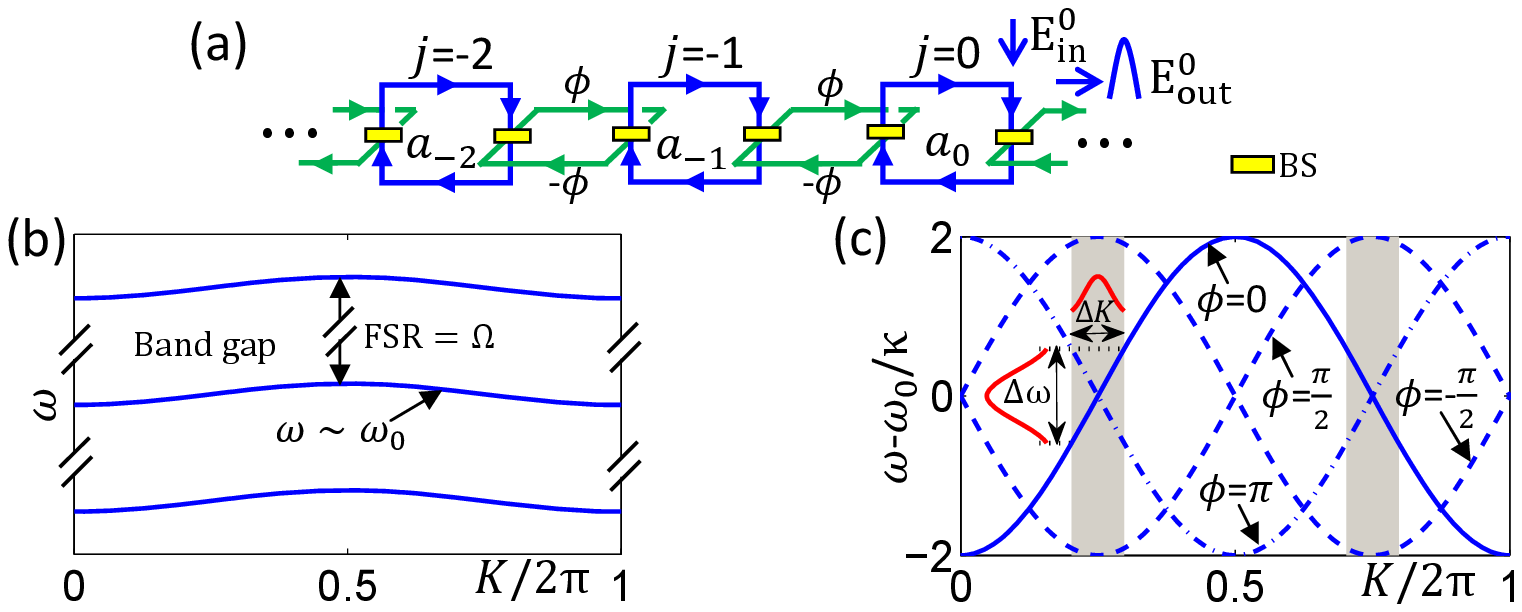}
\caption{\textbf{Spectrum of the proposed all-optical photonic devices in
Fig. \protect\ref{fig:system-1}.} (a) The equivalent circuit in the OAM
space, with $a_{j}$ the field operator for OAM mode $jM$ and $\protect\phi $
the phase imbalance between the two arms of the auxiliary cavity. (b)
Spectrum bands of the system separated by the FSR of the cavity. (c)
Dependence of the spectrum on $\protect\phi $. Momentum and frequency ranges
covered by the signal are marked by the red pulses.}
\label{fig:band-1}
\end{figure}

Interestingly, the single main cavity system in Fig. \ref{fig:system-1} is
conceptually equivalent to a 1D array of coupled optical resonators \cite%
{luo2015quantum}, which makes our scheme much simpler than previous coupled
cavity based quantum devices that contain more than 100 cavity units each
consisting of several carefully coupled and tuned cavities \cite%
{yanik2004stopping}. This mapping is illustrated in Fig. \ref{fig:band-1}
(a), where the $j$-th state with an OAM number $l=jM$ is associated with the
position index of a cavity. In the weak coupling limit between auxiliary and
main cavities, the Hamiltonian for the system in Fig. \ref{fig:band-1} (a)
can be written as \cite{hafezi2011robust}
\begin{equation}
H=\kappa \sum\nolimits_{j}(e^{-i\phi }a_{j}^{\dagger }a_{j+1}+h.c.)+\omega
_{0}\sum\nolimits_{j}a_{j}^{\dagger }a_{j}  \label{eq:H}
\end{equation}%
in the OAM space, where $a_{j}$ is the annihilation operator of the cavity
photon of OAM mode $jM$, $\phi $ is the phase imbalance between the two arms
of the auxiliary cavity, and $\omega _{0}$ is the resonant frequency of the
main cavity. The tunneling rate between OAM modes $\kappa =\Omega _{0}\alpha
(1+\alpha )/2\pi $, where $\alpha =|r_{\text{B}}|^{2}/\left(
1+|t_{B}|^{2}\right) $, $r_{B}$ and $t_{B}$ are the reflectivity and
transmissivity of the coupling beam splitters, $\Omega _{0}=2\pi c/L$ is the
free spectral range (FSR) of the cavity, $L$ is the total length of the
cavity optical path, and $c$ is the speed of light.

For the critical input and output channels of the photonic device, we
introduce a low-reflectivity pinhole at the center of the input/output
mirror as shown in Fig. \ref{fig:system-1} (b), which can be implemented
using, for instance, graded coating \cite{piegari1996coatings}. This is a
widely used technique \cite{barreiro2008beating, wang2012terabit} to
differentiate $l=0$ mode from others since $l=0$ is the only mode with a
high intensity at the beam center \cite{yariv2007photonics, yao2011orbital}.
The rotationally symmetric pinhole does not affect the OAM number
of the cavity modes, and it introduces a large loss rate for the $l=0$ and
low OAM modes, since they can leak out of the cavity via the pinhole and
couple to outside modes. In contrast, higher OAM modes, whose field
distribution has negligible overlap with the pinhole, are hardly disturbed,
just like a $l=0$ mode is not affected by the finite aperture of the mirror
in a cavity (without a pinhole) though its wave front is infinite in theory
(see Supplementary Materials). With proper choice of the step index $M$,
only $j=0$ mode in the cavity couples to the input/output field and has a
significant rate of loss.

When photons propagate in all-optical devices, the time evolution of optical
modes (in Heisenberg picture) is described by \cite{walls2008quantum}%
\begin{eqnarray}
\frac{d}{dt}a_{j}(t) &=&-i\omega _{0}a_{j}(t)+i\kappa \lbrack e^{-i\phi
}a_{j+1}(t)+e^{i\phi }a_{j-1}(t)]  \notag  \label{eq:coupled-m2} \\
&&-\frac{\gamma _{j}}{2}a_{j}(t)+\delta _{j,0}\sqrt{\bar{\gamma}}\hat{E}_{%
\text{in}}^{0}(t),
\end{eqnarray}%
where $a_{j}(t)$ is the time-dependent field operator of OAM mode $jM$, $%
\gamma _{j}$ is its loss rate, and $\hat{E}_{\text{in}}^{0}(t)$ is the input
field operator which couples to the $l=0$ mode in the cavity at a rate $%
\sqrt{\bar{\gamma}}$ determined by the reflectivity of the input/output
pinhole. The input field operator is given by $\hat{E}_{\text{in}}^{0}(t)=%
\frac{1}{\sqrt{2\pi }}\int d\omega b_{\omega }(t_{0})e^{-i\omega (t-t_{0})}$%
, with $t_{0}\rightarrow -\infty $ and $b_{\omega }(t_{0})$ being the
annihilation operator of the input photon with frequency $\omega $. The
proposed devices work for both quantum single-photon and classical coherent
state input pulses (see Supplementary Materials), since the dynamics of our
system is characterized by the linear equation of photon operators (see Eq. %
\ref{eq:coupled-m2}).

The dispersion spectrum of the Hamiltonian (\ref{eq:H}) is%
\begin{equation}
\omega -\omega _{0}=-2\kappa \cos (K-\phi ),  \label{eq:disp}
\end{equation}%
where $\omega $ is the system's eigenfrequency, and $K$ is the Bloch wave
number. Clearly the dispersion relation [see Figs. \ref{fig:band-1} (b) and %
\ref{fig:band-1} (c)] and the propagating group velocity $\frac{\partial
\omega }{\partial K}$ in the OAM space can be manipulated by simply tuning
the phase imbalance $\phi $ in experiments. Such tunability make it possible
to realize important optical devices such as quantum memory and optical
filter.

\subsection{Quantum memory in synthetic OAM
lattices}

Quantum memory is a key element in many quantum information protocols \cite%
{kimble2008quantum, lvovsky2009optical}. Since information is encoded in
photons in a quantum communication network, any non-optical element,
such as atomic ensemble \cite{simon2010quantum}, requires transferring of
information from and back to photons, which complicates the operation of the
quantum memory and lowers its efficiency. Furthermore, only a very limited
number of elements are suitable for atomic-ensemble-based quantum memory,
and the frequency range is restricted to available atomic transitions \cite%
{simon2010quantum}. An all-optical quantum memory eliminates the need to
transfer information between different physical media and can in principle
lead to simplified operation and improved efficiency. However, existing
schemes for all-optical quantum memory based on coupled optical resonators
\cite{yanik2004stopping, yanik2004stopping2} or modulation of index of
refraction \cite{baba2008slow, baba2008large} have their own difficulties
for fabricating large numbers of identical optical cavities or homogeneously
tuning the index of optical materials.

An all-optical quantum memory based on slowing/stopping light through photon
interference in the OAM lattice can overcome those difficulties of existing
schemes and offer compelling advantages. The photon propagation is now
slowed down in the OAM-enabled synthetic lattices by tuning the phase $\phi $%
, which is much simpler and more reliable compared to simultaneous and
precise tuning of hundreds of cavities for quantum memory based on coupled
resonators in real dimension \cite{yanik2004stopping}. The major operation
procedure for the quantum memory consists of three steps by controlling the
phase imbalance $\phi $: \textit{i)} writing the input signal into the
memory by coupling to the $l=0$ mode in the cavity through the input
pinhole; \textit{ii)} letting the signal in the cavity propagate to certain
high $l\neq 0$ modes and storing it there for a desired storage time;
\textit{iii)} making the signal propagate back to the $l=0$ mode for
read-out by coupling to the output through the same pinhole for write-in.

\begin{figure}[tbp]
\includegraphics[width=1.0\linewidth]{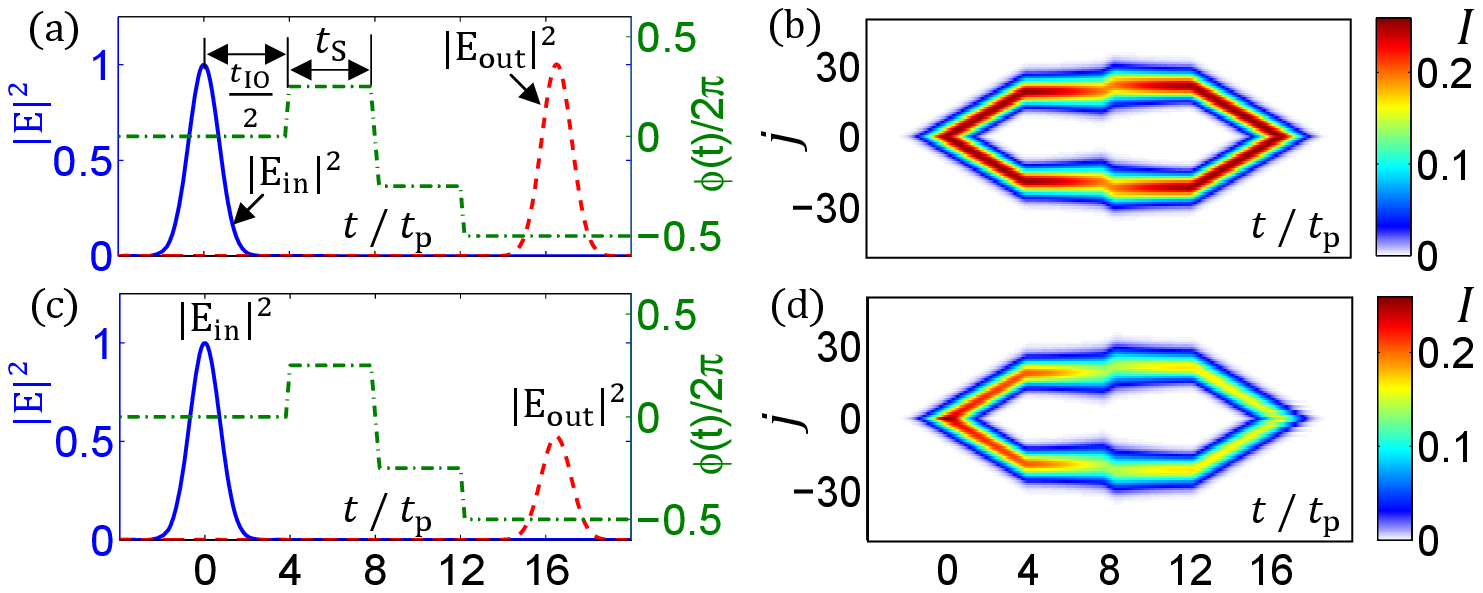}
\caption{\textbf{Time evolution of the optical signal in the OAM-based
quantum memory.} (a) Control sequence for the phase imbalance in the
auxiliary cavity and calculated input and output power normalized to the
maximum intensity of the input pulse, assuming a Gaussian profile for the
input pulse and no loss for all cavity modes except that due to the coupling
to the input signal. $E_{\text{in}}^{0}=\exp (-\frac{t^{2}}{2t_{\text{p}}^{2}%
}-i\protect\omega _{0}t)$ with $t_{\text{p}}=2.5\protect\kappa ^{-1}$ and $%
\protect\gamma _{j}=\protect\delta _{j,0}4\protect\kappa $. (b) The
corresponding evolution of the optical signal's distribution in the OAM
lattice. (c) The same as in (a), except that losses of the cavity modes are
taken into account by assuming $\protect\gamma _{j}=\protect\delta _{j,0}4%
\protect\kappa +0.2\protect\kappa e^{-|j|}+0.01\protect\kappa $ (see
Supplementary Materials). (d) The corresponding evolution in the OAM lattice.
}
\label{fig:memory}
\end{figure}

For a proof-of-principle illustration of our OAM-based quantum memory, we
first ignore the loss of all $l\neq 0$ modes and assume $\gamma _{j}=\delta
_{0j}\bar{\gamma}$. As shown in Figs. \ref{fig:memory} (a) and (b), if we
design the system such that $\bar{\gamma}=4\kappa $, the incoming signal
pulse is absorbed into the cavity with an efficiency of 100\% (see
Supplementary Materials). In order to store pulses significantly shorter
than the write-in time $t_{\text{IO}}$, the usable memory bandwidth $2\kappa
$ \cite{poon2004designing} should satisfy the condition $2\kappa t_{\text{IO}%
}\gtrsim 12\pi $ \cite{simon2007quantum}. Once in the cavity, all frequency
components of the signal pulse start to propagate to $l\neq 0$ modes. For a
long storage time, the signal may propagate to high OAM numbers. Though
there is no theoretical upper limit for the OAM number, in reality it is
limited by practical factors such as the aperture size of the optical
elements in the cavity.

In order to limit the OAM number, we slow down the propagation of the signal
pulse in the OAM space by tuning the phase imbalance $\phi $. As illustrated
in Figs. \ref{fig:memory} (a) and (b), $\phi $ is set to 0 in the write-in
process. When the signal pulse enters the cavity completely after a write-in
time $t_{\text{IO}}$, its peak travels (in the OAM space) approximately at a
group velocity $v_{g}=\frac{\partial \omega }{\partial K}|_{K=\pm \pi
/2}=\pm 2\kappa $. We then change the phase to $\phi =\pi /2$ adiabatically
compared with the bandgap of the system approximately given by the FSR [Fig. %
\ref{fig:band-1} (b)]. The modulation of $\phi $ preserves the system's
translational symmetry in the OAM lattice, and thus the Bloch wave vectors
of the signal is conserved. At $\phi =\pi /2$, $v_{g}$ becomes 0 [see Fig. %
\ref{fig:band-1} (c)], and the pulse stops propagating in the OAM lattice as
shown in Fig. \ref{fig:memory} (b).

Meanwhile, the pulse starts to expand in the OAM lattice due to the
dispersion of the spectrum, which causes distortion in the temporal profile
of the signal. To correct this distortion and restore the signal to its
original shape for read-out, we tune $\phi $ to $-\pi /2$ and keep its value
at $-\pi /2$ for the same amount of time $t_{\text{S}}$ for which $\phi $
was set to $\pi /2$. Finally, we tune $\phi $ to $-\pi $. As shown in Figs. %
\ref{fig:memory} (a) and (b), after another period of time equal to the
write-in time $t_{\text{IO}}$, the above phase echo procedure not only
returns all frequency components of the signal to the $l=0$ mode, but also
corrects any distortion accumulated in the first half of the process. The
pulse can be read out with an efficiency of 100\% under the condition $\bar{%
\gamma}=4\kappa $ and the total storage time $\tau \simeq t_{\text{IO}}+2t_{%
\text{S}}$. To ensure full emission, it is required that $l_{\text{max}%
}/M\gtrsim 2\kappa t_{\text{IO}}\gtrsim 12\pi $, with $l_{\text{max}}$ the
maximum OAM state that the cavity can support.

\begin{figure}[tbp]
\includegraphics[width=1.0\linewidth]{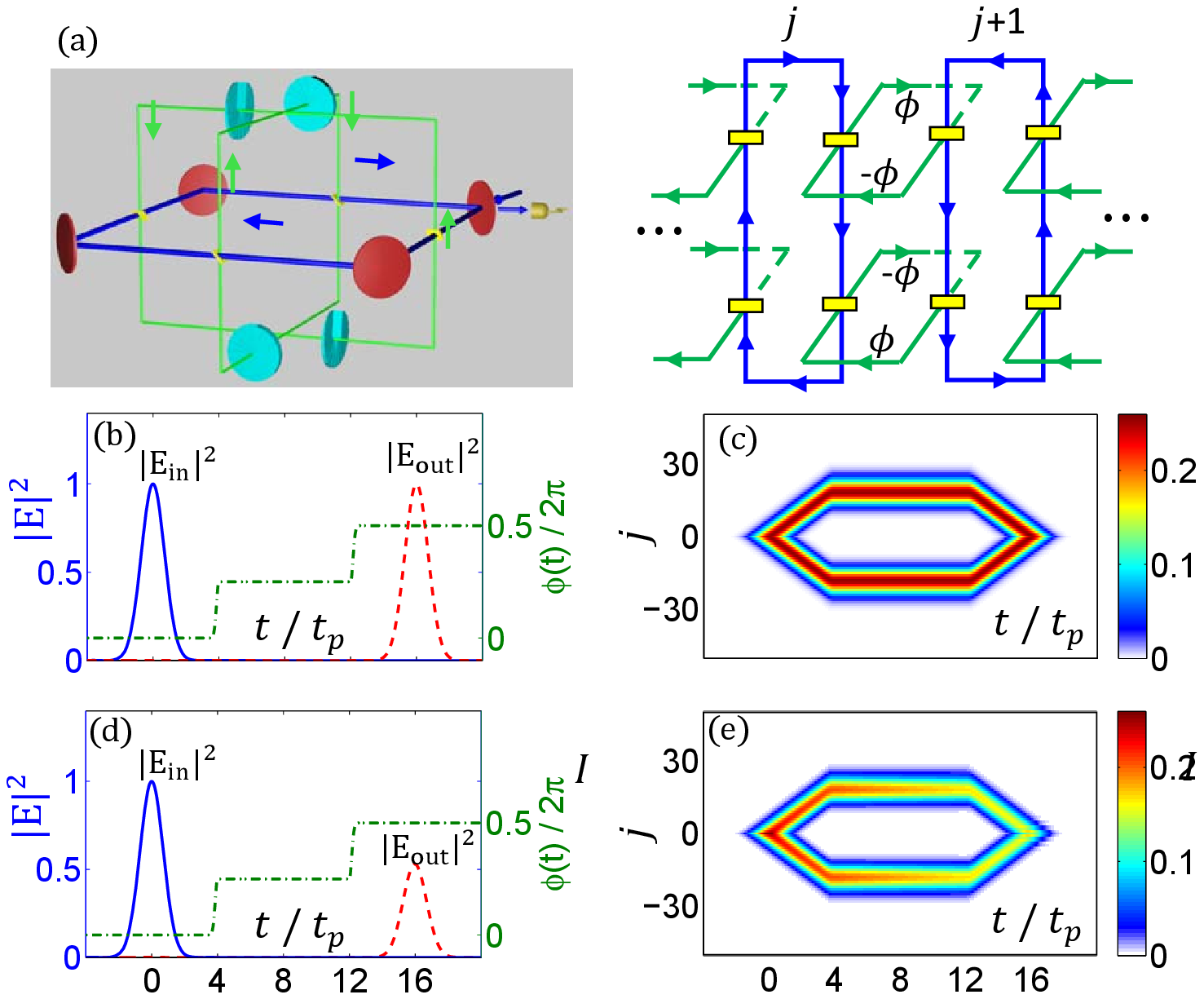}
\caption{\textbf{Modified quantum memory allowing on-demand recall of the
optical signal pulse.} (a) The system with two auxiliary cavities and its
equivalent optical circuit in the OAM lattice. The phase imbalances of the
two auxiliary cavities are opposite to each other. (b) to (e) are the same
as Fig. 4 (a) to (d) except with two auxiliary cavities and slightly
different control sequence for the phase imbalance $\protect\phi $.}
\label{fig:memory_on_demand}
\end{figure}

Although the storage time of our OAM-based quantum memory is controllable,
it is preset. We can freeze the photon signal in the OAM lattice and enable
its on-demand recall by slightly modifying our design from the device in
Fig. \ref{fig:system-1}. The corresponding circuits are shown in Fig. \ref%
{fig:memory_on_demand} (a), which use two auxiliary cavities with the same
coupling strength $\kappa /2$ and opposite phase imbalances $\pm \phi $.
Because of the interference between the two auxiliary cavities, the
dispersion relation of the system becomes
\begin{equation}
\omega -\omega _{0}=-2\kappa \cos \phi \cos K.  \label{eq:disp_on}
\end{equation}%
The group velocities of the pulse peaks at $K=\pm \pi /2$ become $v_{g}=\pm
2\kappa \cos \phi $. Once the input signal is absorbed into the cavity, we
can stop the pulse's propagation and dispersion in the OAM lattice
completely by adiabatically changing $\phi $ from 0 to $\pi /2$, which
compresses the bandwidth to 0 because transitions between OAM modes via the
two auxiliary cavities cancel each other. As shown in Fig. \ref%
{fig:memory_on_demand}, the optical signal and its distribution in the OAM
lattice can then be kept for an arbitrary and indefinite amount of time
until it needs to be read-out by changing $\phi $ from $\pi /2$ to $\pi $.
This allows the on-demand recall of the photon signal and random access to
the quantum information that it carries. The storage fidelity of a
single-photon pulse, which is defined as the wave-packet overlap between
input and output conditional on the re-emission of a photon \cite%
{simon2010quantum}, is close to 1. Thus it is possible to realize perfect
write-in, storage, and on-demand read-out of an optical signal using only a
limited number of OAM states sufficient for the signal pulse to be absorbed
into the cavity.

In reality, all OAM modes are lossy due to factors such as intrinsic loss of
the optical elements and leakage of $l\neq 0$ modes via the input-output
pinhole. It is demonstrated that (see Supplementary Materials), our
OAM-based quantum memory still functions as expected without wave-packet
distortion in the presence of imperfections, though the efficiency is
reduced, as shown in Figs. \ref{fig:memory} (c), (d) and Figs. \ref%
{fig:memory_on_demand} (d) and (e).

For the estimation of experimental parameters, we assume that the cavity is
realized using four curved mirrors each with a focal length $F$ on the order
of centimeters, a typical value for discrete optical elements.
Since the separation between the mirrors is $2F$, the total length is about
tens of centimeters for the optical path, which gives a FSR ($\Omega _{0}$)
of $2\pi \times 0.5$GHz - $2\pi \times 1.0$GHz. By choosing a proper
reflectivity $r_{B}^{2}\sim 0.25$ for the beam splitters, we estimate that
the total bandwidth $4\kappa $ is about $2\pi \times 50$MHz - $2\pi \times
100$MHz. Therefore, the quantum memory can store short pulses with a
temporal duration of tens of nanoseconds. The bandwidth can be further
improved by using a smaller focal length $F$. The required modulation time
for the phase imbalance $\phi $ is also on the order of tens of nanoseconds,
consistent with the modulation speed of current electro-optic devices \cite%
{chuang1995physics, yariv2007photonics}. With a photon loss rate of the
order of MHz, the storage time is about $1\mu $s (see Supplementary
Materials). As a comparison, the storage time of ideally identical
coupled-micro-resonator based memory is limited below $0.1\mu $s due to the
large photon losses of the micro-resonators (about tens of MHz) \cite%
{yanik2004stopping}.

\begin{figure}[tbp]
\includegraphics[width=1.0\linewidth]{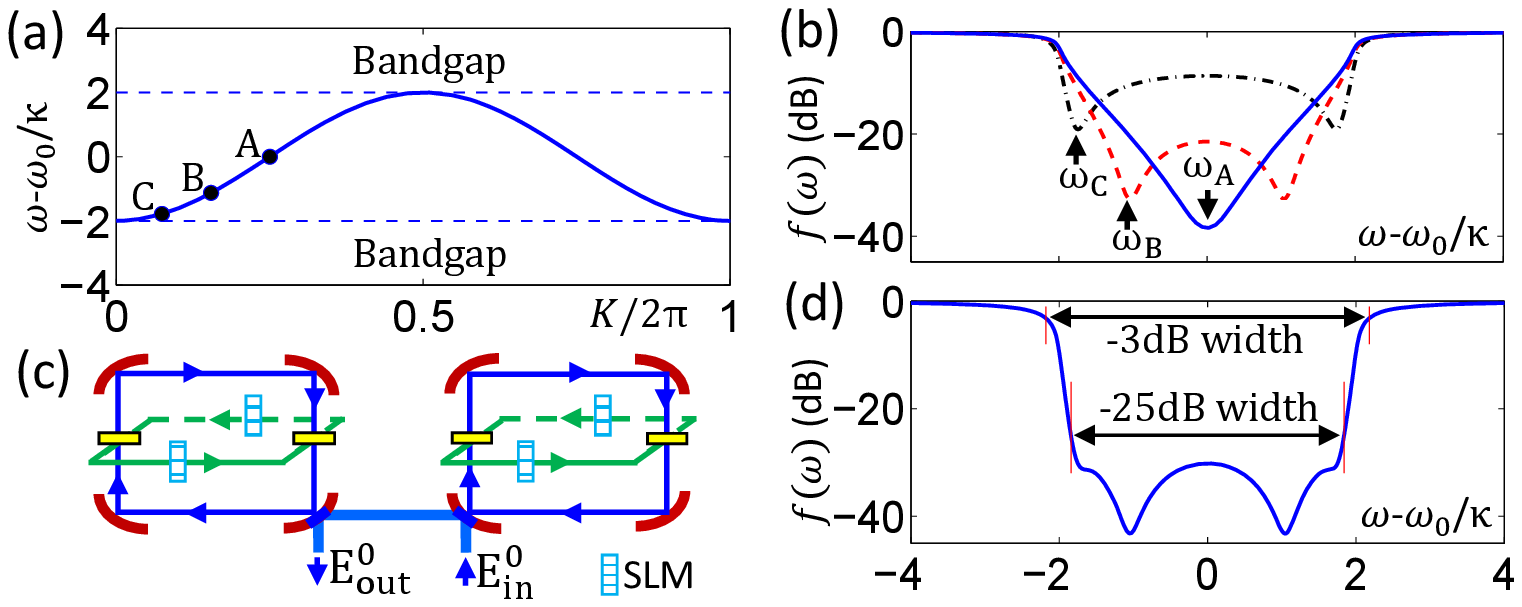}
\caption{\textbf{Design of a stopband optical filter based on the OAM of
photons.} (a) Spectrum band of the device in Fig. \protect\ref{fig:system-1}
with 3 marked frequencies $\protect\omega _{A}=\protect\omega _{0}$, $%
\protect\omega _{B}\simeq \protect\omega _{0}-1.1\protect\kappa $, and $%
\protect\omega _{C}\simeq \protect\omega _{0}-1.8\protect\kappa $. The
corresponding group velocities are $v_{g}(\protect\omega _{A})=2\protect%
\kappa $, $v_{g}(\protect\omega _{B})=1.65\protect\kappa $, and $v_{g}(%
\protect\omega _{C})=0.9\protect\kappa $. (b) The filter function $f(\protect%
\omega )=|E_{\text{out}}^{0}|^{2}/|E_{\text{in}}^{0}|^{2}$ (see
Supplementary Materials) of the device in Fig. \protect\ref{fig:system-1}
when its maximum absorption frequency is designed to be $\protect\omega _{A}$%
, $\protect\omega _{B}$, and $\protect\omega _{C}$ respectively. A maximum
absorption frequency closer to the band edge results in a steeper skirt
slope but poorer in-band rejection ratio. (c) A two-cavity design with their
maximum absorption frequencies chosen to be $\protect\omega _{C}$ and $%
\protect\omega _{B}$ respectively. (d) The filter function for the design in
(c). In (b) and (d), $\protect\gamma _{j}=\protect\delta _{j,0}\bar{\protect%
\gamma}+0.1\protect\kappa e^{-|j|}+0.1\protect\kappa $ . }
\label{fig:filter}
\end{figure}

\subsection{OAM-enabled optical filter}

We can build upon our ideas to envision further interesting and valuable
applications. One such example is high skirt-slope optical filters which are
crucial in many fields such as quantum information \cite%
{palittapongarnpim2012note, neergaard2006generation, macrae2012tomography},
high-density wavelength-division-multiplexing networking, and optical signal
processing \cite{ilchenko2006optical, minasian2001photonics, little2004very}%
. For good selectivity, the filter function should ideally have a narrow
bandwidth and a steep skirt slope at the edge of the stopband. It is then
critical to improve the shape factor which is often evaluated by the ratio
of the stopband width at -25dB and -3dB \cite{minasian2001photonics,
little2004very}. Conventionally, this is usually achieved by coupling many
carefully designed cavities to obtain a high-order filter \cite%
{little2004very}. Because of inevitable errors in fabrication and tuning,
the number of cavities that can be reliably coupled in practice is quite
limited. Consequently, it is very challenging to realize high shape factors
in an optical filter based on many coupled cavities.

It is possible to achieve an optical filter with very high shape factors
using the band spectrum Eq. (\ref{eq:disp}) generated by the photon
interference in the OAM lattice in Fig. \ref{fig:filter}. The filter
characteristics can be obtained by analyzing the wave propagation in the
coupled many-cavity system in Fig. \ref{fig:band-1} (a) which is a
conceptual equivalent to our OAM-based device. However, a much more
intuitive understanding based on the system spectrum is possible which can
greatly facilitate the design of the filter to obtain desired properties. In
the device in Fig. \ref{fig:system-1}, when an $l=0$ signal is fed to the
input/output port, all frequency components in the bandgaps cannot enter the
cavity and are reflected. The cavity absorbs in-band frequency components
with an efficiency dependent on the coupling rate $\sqrt{\bar{\gamma}}$ and
the group velocity $v_{g}(\omega )$ of the $l=0$ cavity mode in the OAM
lattice. The maximum absorption occurs at the frequency $\omega _{\text{m}}$
determined by $2|v_{g}(\omega _{m})|=\bar{\gamma}$ (see Supplementary
Materials). If we choose the reflectivities of the coupling beam splitter
and input/output pinhole appropriately such that $\omega _{m}$ is very close
to the cavity's band edge $\omega _{\text{e}}=\omega _{0}\pm 2\kappa $, the
cavity changes from being totally reflective to being strongly absorptive to
the incident light over a narrow frequency range $|\omega _{\text{m}}-\omega
_{\text{e}}|$. This results in a desired steep skirt slope as shown in Fig. %
\ref{fig:filter} (b). Unfortunately, because of the frequency dependence of
the group velocity, for such a choice of $\omega _{m}$ the absorption is
poor at the center of the stopband, leading to an insufficient in-band
rejection ratio which manifests as the hump at the bottom of the filter
function in Fig. \ref{fig:filter} (b).

In order to overcome this difficulty, we use the two-cavity design in Fig. %
\ref{fig:filter} (c). While the maximum absorption frequency of the first
cavity is still chosen to be close to the band edge, that of the second
cavity is chosen closer to the center of the stopband to suppress the hump
in Fig. \ref{fig:filter} (b). Such a design results in a narrow and deep
stopband with sharp edges, which is ideal for optical filters.

Since the input and output fields are both in the $l=0$ mode, the filter
function for our filter is calculated by \cite{luo2015quantum}
\begin{equation}
f(\omega)=|\langle l=0|1+G|l=0\rangle|^2
\end{equation}
with
\begin{equation}
G=\frac{-i\bar{\gamma}}{\omega-\sum_K\omega_{K}|K\rangle\langle K|+i\sum_l%
\frac{\gamma_l}{2}|l\rangle\langle l|},
\end{equation}
where $|l\rangle$ is an OAM state, $|K\rangle$ is a Bloch state in the OAM
space with frequency $\omega_{K}=\omega_0-2\kappa\cos K$. As plotted in Fig. %
\ref{fig:filter} (d), a shape factor of 0.85 can be realized with just
moderate SLM and cavity efficiencies (see Supplementary Materials), which is
noticeably higher than current technologies that are limited by the number
of high-Q cavities that can be reliably coupled in practice \cite%
{little2004very, savchenkov2009narrowband, rasras2009demonstration}.

\section{Discussion}

Imperfections such as photon losses will degrade the performance of the
optical devices. Intrinsic loss due to the finite finesse of the cavity can
be very low as long as high quality cavities are used \cite%
{nagorny2003collective}. Photon losses can also be introduced by phase
modulators due to absorption by their optical media and SLMs
because of their limited resolution and fabrication error. Such losses can
be made very low \cite{leidinger2015highly, oemrawsingh2005experimental,
marrucci2011spin, raut2011anti} (see Supplementary Materials), and further
reduced by the fact that the auxiliary cavity is designed using destructive
interference with very few photons in it. The low reflectivity pinhole manifests
as an photon loss that decreases rapidly with the OAM number (see
Supplementary Materials). For the quantum memory, the storage time (for a
fixed storage efficiency) decreases rapidly with the increase of photon
losses. Photon losses due to the phase modulators and SLMs are the limiting
factors for the storage time, since their effect is persisting, even during
the storage phase when the signal is frozen at large OAM states. We find
that for the optical filter, the shape factor and stop bandwidth are less
sensitive to imperfections of these optical elements, as confirmed by our
numerical simulation.

Even with the limitations posed by these practical considerations, the
OAM-based quantum memory has a few attractive characteristics and
note-worthy advantages that are not available in existing schemes. Not only
is the system very simple with just a single main cavity and thus completely
realizable with conventional optical technology, but also the operating
wavelength can be chosen at will, a significant edge in situations where no
atomic systems with the desired transition frequency are available. It is
also not limited by the technical challenge to fabricate and tune a large
number of identical optical cavities \cite{yanik2004stopping}. The bandwidth
of the quantum memory and its storage time are limited by the size of the
cavity and loss of the SLMs and phase modulators, instead of the
delay-bandwidth product of the system or other intrinsic factors. Finally,
polarization independent optical elements can be used so that information
encoded in both temporal wave-packet and polarization can be recovered with
fidelity close to 1.

In conclusion, we propose a conceptually novel route for engineering
all-optical photonic devices based photon propagation and interference in
synthetic lattices. We demonstrate this new designing principle by showing
that two powerful devices, quantum memory and optical filter, can be
realized utilizing photon OAM-based synthetic lattices. The proposed route
may inspire new and simple designs for many other photonic devices (e.g.,
multi-channel optical router, etc.), and open a completely new avenue for
photonic technology and applications.

\section{Acknowledgements}

This work is funded by NNSFC (Grant Nos. 11574294, 61490711),
NKRDP (Grant Nos.2016YFA0301700 and 2016YFA0302700) and the "Strategic Priority
Research Program(B)" of the CAS (Grant No.XDB01030200). CZ is supported by ARO
(W911NF-12-1-0334) and NSF (PHY-1505496).

\end{document}